\documentstyle[pra,aps,epsfig]{revtex}
\tighten

\newcommand{\ket}[1]{\left | #1 \right \rangle}
\newcommand{\bra}[1]{\left \langle #1 \right |}

\newcommand{\proj}[1]{\ket{#1} \bra{#1}}

\begin{document}
\draft

\title{Fidelity and the communication of quantum information}
\author{Stephen M. Barnett$^1$, Claire R. Gilson$^2$ and Masahide 
Sasaki$^{3,4}$}
\address{${}^1$
    Department of Physics and Applied Physics, University of
    Strathclyde,
    Glasgow G4 0NG, Scotland}
\address{${}^2$
    Department of Mathematics, University of Glasgow,
    Glasgow G12 8QW, Scotland}
\address{${}^3$
    Communications Research Laboratory,
    Koganei, Tokyo 184-8795, Japan}
\address{${}^4$CREST, Japan Science and Technology}

%\date{30 Jan 2001}

\maketitle

\begin{abstract}
We compare and contrast the error probability and fidelity
as measures of the quality of the receiver's measurement
strategy for a quantum communications system.  The error
probability is a measure of the ability to retrieve {\it classical}
information and the fidelity measures the retrieval of
{\it quantum} information.  We present the optimal measurement
strategies for maximising the fidelity given a source that
encodes information on the symmetric qubit-states.

\end{abstract}


%\begin{multicols}{2}

\section{Introduction}\label{introduction}

The principles governing the communication of information by
a quantum channel are now well-known
\cite{Helstrom_QDET,Holevo_Book,Hirota_Book}.
The transmiting party (Alice)
selects from a set of signal states $\ket{\psi_j}$ and uses a
string of these to encode her message.  These states are known
to the receiving party (Bob) who also knows the {\it a priori} 
probabilities $p_j$ for selection of each of the signal states.
Bob's problem is to decide upon an optimal detection strategy.
His choice of strategy will depend on the way in which the
information he receives is to be used.  In mathematical terms,
Bob must choose a strategy so as to extremise some function
of his measurement outcomes and commonly occurring examples
are the minimum error probability or minimum Bayes cost
\cite{Helstrom_QDET,Holevo_Book,Hirota_Book,Yuen_75,Ban_97}
and the accessible information
\cite{Davies_78,Fuchs_96,Ban_97a,Osaki_98,Sasaki_99}.  These
quantities determine the quality of Bob's strategy for recovering
the classical information associated with Alice's selection of
the transmitted state.

In this paper we will be concerned with a different measure
of Bob's detection strategy.  This quantity, which we refer
to as the fidelity, determines Bob's ability to access the
{\it quantum} information contained in Alice's signal.  The
fidelity depends on Bob's choice of measurement strategy
and also on his subsequent selection of a new quantum state.
The extent to which the selected state matches that chosen
by Alice will determine Bob's ability to reconstruct the
selected quantum state.  We will introduce the fidelity and
compare its properties with those of the more familiar
error probability in the following section.  At this stage,
we can motivate our idea by considering the familiar
problem of eavesdropping in quantum key distribution
\cite{Phoenix_95}.  The error probability and fidelity
relate, in this case, to the two principal factors in assessing
any eavesdropping strategy.  The error probability is
simply the probability that the eavesdropper will fail to learn
the state selected by Alice, while the fidelity is the
probability that the state selected by the eavesdropper
for transmission to Bob will appear to Bob as the
state selected by Alice.  In this way, error probability
is related to the security of the classical information
encoded by Alice and the fidelity is related to the
likelihood of escaping detection \cite{foot1}.

We have not been able to find general criteria for maximising
the fidelity.  This maximum fidelity was introduced by Fuchs
\cite{Fuchs_00} who referred to it as the accessible fidelity.
For a special class of qubit-states known as the
symmetric states, however, we have been able to derive the
strategy that maximises the fidelity.  The measurement part
of the optimal strategy is not unique, but includes the strategy
that also minimises the error probability \cite{Ban_97}.

\section{Fidelity and error probability}\label{fidelity}

In a quantum communications channel, Bob's problem is
to distinguish between the set of possible signal states,
$\ket{\psi_j} (j=1,...M)$, that Alice may have sent.  He
does this by performing a measurement the results of
which are associated with the POM elements
\cite{Helstrom_QDET,Peres_Book} $\hat\pi_k$.
There is, of course, no particular reason for the
number of possible measurement outcomes to
equal {\it M}, the number of possible signal states.
The probability that Bob observes the result `{\it k}'
given that Alice selected the state $\ket{\psi_j}$
is
\begin{equation}
P(k\vert j)=\langle\psi_j\vert\hat\pi_k\vert\psi_j\rangle.
\label{cond_prob}
\end{equation}
If Bob wishes to determine the signal state then the
probability that he will do so correctly is
\begin{equation}
P_{\rm c}=\sum_{j=1}^M P(j\vert j)p_j
=\sum_{j=1}^M \langle\psi_j\vert\hat\pi_j\vert\psi_j\rangle p_j.
\label{prob_correct_decision}
\end{equation}
This quantity is a measure of the success of Bob's
strategy at recovering Alice's (classical) choice
of signal state.  The error probability is simply $1 - P_{\rm c}$:
\begin{equation}
P_{\rm e}=1-\sum_{j=1}^M
                 \langle\psi_j\vert\hat\pi_j\vert\psi_j\rangle p_j.
\label{error_prob}
\end{equation}
Necessary and sufficient conditions are known for
minimising $P_{\rm e}$ (or maximising $P_{\rm c}$)
\cite{Helstrom_QDET,Holevo_Book,Hirota_Book,Yuen_75}
although very few explicit examples of the required
POM elements have been given.  Some of these minimum
error POMs have recently been implemented
optically \cite{Barnett_97,Clarke_01,Fujiwara_01}.

The fidelity is more closely related to the retrieval of the
quantum information `$\ket{\psi_j}$'.  As a physical picture,
consider Bob to be operating some relay station in a
communications channel.  He must measure the signal and
then, on the basis of his measurement, he selects a state
to retransmit.  The fidelity is then a measure of how well
the selected state matches the original signal state selected
by Alice.  We can see this by considering one of the possible
sequence of events.  Let us suppose that Alice has sent the
signal state $\ket{\psi_j}$ and that Bob measurement has
given the result `{\it k}' corresponding to the POM element
$\hat\pi_k$.  He then selects a state, $\ket{\phi_k}$,
that depends on the measurement result, for retransmission.
The simplest question that we can ask, to assess the
retransmitted state, is ``is this state $\ket{\psi_j}$ ?".
The probability that this question will be answered
in the affirmative is just the modulus squared
overlap of the signal state and the retransmitted state,
$\vert\langle\psi_j\vert\phi_k\rangle\vert^2$.
The {\it a priori} probability that the retransmitted
state will pass this test is the fidelity
\begin{equation}
F=\sum_{j=1}^M\sum_k
     \vert\langle\psi_j\vert\phi_k\rangle\vert^2
     \langle\psi_j\vert\hat\pi_k\vert\psi_j\rangle p_j.
\label{F}
\end{equation}
This quantity determines the quality of the
measurement-retransmission strategy adopted by Bob.  The strategy
adopted by Bob depends on both his choice of measurement
(associated with the POM elements $\hat\pi_k$)
and the selection of the associated retransmission states
($\ket{\phi_k}$).  A large value of {\it F} corresponds to a
good strategy while a smaller value indicates a less good one.
The strategy that best extracts the quantum information
will be the one that gives the maximum fidelity.
The general principles governing the maximum fidelity
are unknown to us although the maximum fidelity,
the associated measurement and retransmission states
have been derived for a special case \cite{Fuchs_00}.
We will present strategies for maximising the fidelity
for a wider set of possible signal states
(the symmetric qubit-states) in section IV.

\section{Symmetric states}\label{Symmetric}

The symmetric states were introduced for the problems of
state discrimination by Ban {\it et. al.} \cite{Ban_97}.
These states, $\ket{\psi_j}$, are generated from a single
state, $\ket{\psi_1}$, by the action of a unitary operator
$\hat V$:
\begin{equation}
\ket{\psi_j}=\hat V^{j-1}\ket{\psi_1}.
\label{psi_j}
\end{equation}
These {\it M} states are said to be symmetrical if they
are {\it a priori} equally likely to have been selected
and
\begin{equation}
\hat V^M=\hat I
\label{VM=1}
\end{equation}
so that $\ket{\psi_{j+M}}=\ket{\psi_j}$ \cite{foot3}.

\begin{figure}[htb]
\centerline{\psfig{file=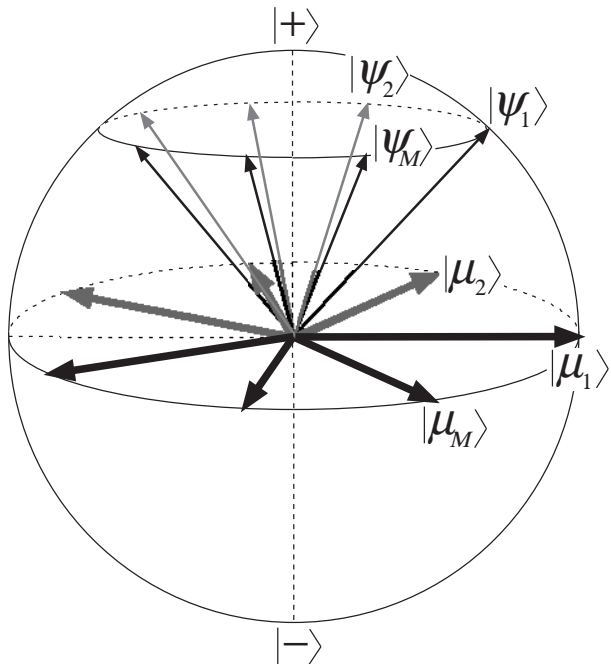,width=8cm}}
\caption{The symmetric set of states $\{\ket{\psi_j}\}$.
The square root measurement has POM elements that
are proportional to projectors onto the states $\{\ket{\mu_j}\}$.
This measurement minimises the average error probability
for distinguishing between the states.}
\label{sig&SRmeas}
\end{figure}

The minimum error probability occurs \cite{Ban_97}
if we adopt the so-called square-root measurement
\cite{Holevo_SubOptMeas78,Hausladen_SubOptMeas95,Hausladen96_coding}
for which the {\it M} POM elements are
\begin{equation}
\hat\pi_k=\hat\Phi^{-1/2}\proj{\psi_k}\hat\Phi^{-1/2},
\label{pi_k}
\end{equation}
where
\begin{equation}
\hat\Phi=\sum_{j=1}^M\proj{\psi_j}.
\label{Phi}
\end{equation}
The resulting minimum error probability is then
\begin{equation}
P_{\rm e}^{\rm min}=1-
\vert\langle\psi_1\vert\hat\Phi^{-1/2}\vert\psi_1\rangle\vert^2.
\label{min_error_prob}
\end{equation}
In this paper we will obtain the maximum fidelity for any
symmetric states of a single qubit.  We can represent
these states in terms of the orthonormal eigenstates,
$\ket\pm$ of the unitary operator
\begin{equation}
\hat V={\rm exp}\left[i{{2\pi}\over{M}}\proj{-}\right].
\label{V}
\end{equation}
This operator clearly satisfies the requirement Eq. (\ref{VM=1})
for a symmetric set of states.  Our {\it M}, equiprobable
symmetric states are
\begin{equation}
\ket{\psi_j}= {\rm cos}\left({\theta\over2}\right)\ket{+}
                    + {\rm exp}\left(i{{2\pi}\over{M}}(j-1)\right)
                       {\rm sin}\left({\theta\over2}\right)\ket{-},
\quad (0\le\theta\le{\pi\over2}).
\label{general_psi_j}
\end{equation}
It is helpful to picture these states on the Bloch sphere
(see Fig. \ref{sig&SRmeas}).
%%%%%%%%%%%%%%%%%%%%
%%%%%%%%%%%%%%%%%%%%
Each of the states is represented by a
point on the surface of the sphere with the polar coordinates,
$\theta$ and $\phi$, corresponding to $\theta$ and $2\pi j/M$
respectively in (Eq. \ref{general_psi_j}).  The symmetric states lie
on a single circle of the Bloch sphere at the latitude
$\pi /2 - \theta$.   For this set of symmetric states
the minimum error probability is obtained by means of
a POM with elements
\begin{equation}
\hat\pi_j={2\over M}\proj{\mu_j},
\label{pi_mu}
\end{equation}
where
\begin{equation}
\ket{\mu_j}= {1\over{\sqrt2}}
                       \left[ \ket{+}
                    + {\rm exp}\left(i{{2\pi}\over{M}}(j-1)\right)
                       \ket{-}\right].
\label{mu}
\end{equation}
These states correspond to points on the equator of the
Bloch sphere at the same longitude ($\phi$ - coordinate)
as the corresponding signal states $\ket{\psi_j}$
(see Fig. \ref{sig&SRmeas}).
The associated minimum error probability  is
\begin{equation}
P_{\rm e}^{\rm min}=
1-{1\over M}\left(1+{\rm sin}\theta\right).
\label{final_min_error_prob}
\end{equation}
As $\theta$ varies between 0 and $\pi /2$ this error probability
varies between $1-1/M$ and $1-2/M$.  These values correspond
to guessing the value `{\it j}' when the states all correspond
to the single ket $\ket{+}$ and the minimum attainable error
probability for symmetric states \cite{Yuen_75,Clarke_01}
which occurs when the
symmetric states lie on the equator of the Bloch sphere.

In the following section we establish the maximum fidelity
attainable for this ensemble of states.

\section{Maximum fidelity}\label{Maximum}

In seeking to maximise the fidelity it is helpful to write
it in the form \cite{Fuchs_00}
\begin{equation}
F=\sum_k \langle\phi_k\vert\hat O_k\vert\phi_k\rangle,
\label{F_max_eigenval}
\end{equation}
where $\hat O_k$ is the Hermitian operator
\begin{equation}
\hat O_k={1\over M}\sum_{j=1}^M
\vert\psi_j\rangle\langle\psi_j\vert
\hat \pi_k
\vert\psi_j\rangle\langle\psi_j\vert.
\label{O_k}
\end{equation}
The selection of the retransmission states $\ket{\phi_k}$
is now straightforward.  The best state to select will be
the eigenstate of $\hat O_k$ having the largest eigenvalue
and the corresponding maximum fidelity is the
sum of the maximum eigenvalues of the operators $\hat O_k$
\cite{Fuchs_00}.

The problem of maximising {\it F} is now simply one of
selecting the POM or POMs that produce the largest
eigenvalue sum.  Naturally, there are constraints
associated with the fact that our POM elements must
be Hermitian, positive-semidefinite and must sum to
the identity.  In seeking the optimal POM, it is sufficient
to consider only rank-one elements correponding to
weighted projectors onto pure states \cite{foot2}.
The (rank-one) POM elements can be written in
the form
\begin{equation}
\hat\pi_k=2w_k
    \left[
              {\rm cos}\left({\theta_k\over2}\right)\ket{+}
           + {\rm e}^{i\phi_k}
              {\rm sin}\left({\theta_k\over2}\right)\ket{-}
    \right]
    \left[
              {\rm cos}\left({\theta_k\over2}\right)\bra{+}
           + {\rm e}^{-i\phi_k}
              {\rm sin}\left({\theta_k\over2}\right)\bra{-}
    \right]
\end{equation}
or, more simply, as the matrix
\begin{equation}
\hat\pi_k=w_k
\left(
\begin{array}{cc}
1+\cos {\theta_k} & {\rm e}^{-i\phi_k}\sin {\theta_k}  \\
{\rm e}^{i\phi_k}\sin {\theta_k} & 1-\cos {\theta_k}
\end{array}
\right),
\end{equation}
where the basis states $\ket +$ and $\ket -$ correspond
to the column vectors $(1, 0)^T$ and $(0, 1)^T$ respectively.
Here,  $w_k$ is a weight factor bounded by
$0 \le w_k \le 1$.  The requirement that the POM
elements should sum to the
identity places restrictions in the allowed values of the
parameters $\theta_k$, $\phi_k$ and $w_k$.  These
take the form:
\begin{equation}
\sum_k w_k=1,
\label{constraint1}
\end{equation}
\begin{equation}
\sum_k w_k \cos {\theta_k}=0,
\label{constraint2}
\end{equation}
\begin{equation}
\sum_k w_k {\rm e}^{i\phi_k}\sin {\theta_k}=0.
\label{constraint3}
\end{equation}

Our first task is to obtain the greater of the two eigenvalues
for each of the operators $\hat O_k$.  Evaluating the sum
in Eq. (\ref{O_k}) and writing the resulting operator in matrix
form gives
\begin{equation}
\hat O_k={w_k\over2}
\left(
\begin{array}{cc}
(1+\cos{\theta})(1+\cos{\theta}\cos{\theta_k})
   &{1\over2}\sin^2{\theta}\sin {\theta_k}({\rm e}^{-i\phi_k}
+\delta_{M,2}{\rm e}^{i\phi_k})  \\
{1\over2}\sin^2{\theta}\sin {\theta_k}({\rm e}^{i\phi_k}
+\delta_{M,2}{\rm e}^{-i\phi_k})
   &(1-\cos{\theta})(1+\cos{\theta}\cos{\theta_k})
\end{array}
\right),
\label{O_k_gen}
\end{equation}
where $\delta_{M,2}$ is the usual Kronecker delta.
We see that this matrix has one of two possible forms,
one if $M>2$ and one if $M=2$.  It is simplest to deal
these two cases separately.

\subsection{Case 1: $M>2$}

If we have more than two signal states then the
operator Eq. (\ref{O_k_gen}) reduces to
\begin{equation}
\hat O_k={w_k\over2}
\left(
\begin{array}{cc}
(1+\cos{\theta})(1+\cos{\theta}\cos{\theta_k})
   &{1\over2}\sin^2{\theta}\sin {\theta_k}{\rm e}^{-i\phi_k}\\
{1\over2}\sin^2{\theta}\sin {\theta_k}{\rm e}^{i\phi_k}
   &(1-\cos{\theta})(1+\cos{\theta}\cos{\theta_k})
\end{array}
\right).
\label{O_k_enroute}
\end{equation}
The two eigenvalues of this matrix are
\begin{equation}
\nu
_\pm(\theta_k)={w_k\over2}
\left\{
1+\cos{\theta}\cos{\theta_k}
\pm
\left[
\cos^2{\theta}
(1+\cos{\theta}\cos{\theta_k})^2
+
{1\over4}\sin^4{\theta}\sin^2{\theta_k}
\right]^{1/2}
\right\},
\end{equation}
so the maximum value of the fidelity has the form
\begin{equation}
F=\sum_k\nu_+(\theta_k)
={1\over2} +{1\over2}\sum_k w_k
\left[
\cos^2{\theta}
(1+\cos{\theta}\cos{\theta_k})^2
+
{1\over4}\sin^4{\theta}\sin^2{\theta_k}
\right]^{1/2},
\label{max_fidelity}
\end{equation}
where we have used Eqs. (\ref{constraint1}) and (\ref{constraint2}).
Our remaining task is to maximise this quantity subject to
the constraints that the operators $\hat \pi_k$
form a POM Eqs. (\ref{constraint1}-\ref{constraint3}).  The natural 
approach to tackling
such constrained extremisation problems is to use
Lagrange's method of undetermined multipliers.
Before performing this maximisation we note that
the maximum fidelity Eq. (\ref{max_fidelity}) does not depend on the
phases $\phi_k$.  This means that the contribution
to the fidelity will be the same for each POM element
having the same value of $\theta_k$.  Hence we can
easily impose the constraint Eq. (\ref{constraint3}) by choosing a
POM with {\it N} elements for each distinct value of
$\theta_k$ satisfying the simpler condition
\begin{equation}
\sum_{l=1}^N w_l(k)
{\rm e}^{i\phi_l}=0,
\end{equation}
where $w_l(k)$ are the weights associated with the
$N$ POM elements for which $\theta=\theta_k$.
Hence we will not impose the constraint Eq. (\ref{constraint3})
in our variational calculation.  In order to
impose the remaining constraints,
Eqs. (\ref{constraint1}) and (\ref{constraint2}),
we introduce the zero-valued quantities
\begin{equation}
G_1=\sum_k w_k -1,
\end{equation}
and
\begin{equation}
G_2=\sum_k w_k \cos{\theta_k}.
\end{equation}
The extrema of the fidelity will be given by the
stationary points of the function
\begin{equation}
H=F + \lambda_1 G_1 + \lambda_2 G_2
\end{equation}
under independent variation of the parameters $\theta_k$
and $w_k$.  Here $\lambda_1$ and
$\lambda_2$ are the undetermined multipliers.

The stationarity condition for variation of
{\it H} with respect to $w_k$ gives
\begin{equation}
{{\partial H}\over{\partial w_k}}
={1\over2}
\left[
\cos^2{\theta}
(1+\cos{\theta}\cos{\theta_k})^2
+
{1\over4}\sin^4{\theta}(1-\cos^2{\theta_k})
\right]^{1\over2}
+\lambda_1 + \lambda_2 \cos{\theta_k}
=0.
\label{dH_by_dw}
\end{equation}
while variation with respect to $\theta_k$ gives
\begin{eqnarray}
{{\partial H}\over{\partial \theta_k}}
&=&-\sin{\theta_k}{w_k\over2}
\left\{
\left[
\cos^2{\theta}
(1+\cos{\theta}\cos{\theta_k})^2
+
{1\over4}\sin^4{\theta}(1-\cos^2{\theta_k})
\right]^{-{1\over2}}
\left[
\cos^3{\theta}
+ \cos{\theta_k}
\left(
\cos^4{\theta}
-{1\over4}\sin^4{\theta}\right)
\right]
+2 \lambda_2
\right\} \nonumber\\
&=&
0.
\label{dH_by_dtheta}
\end{eqnarray}
The possible solutions of Eq. (\ref{dH_by_dtheta}) are (i) $w_k =0$
corresponding to uninteresting zero POM elements,
(ii) ${\rm sin}\theta_k=0$, corresponding to
POM elements that are proportional to
$\ket{+}\bra{+}$ and $\ket{-}\bra{-}$,
and (iii) the function
in curly parentheses is zero.  This final condition
reduces, by use of Eq. (\ref{dH_by_dw}) to
\begin{equation}
\cos{\theta_k}=
{
  {4\lambda_1\lambda_2 -\cos^3{\theta}}
  \over
  {\cos^4{\theta}-{1\over4}\sin^4{\theta}-4\lambda_2^2}
}
=\cos\Theta.
\end{equation}
which has only {\it one} solution.

Remarkably, we can conclude that the strategy for achieving
the maximum fidelity depends only on three
possible values of $\theta_k$, these being
0,  $\pi$ and the, yet to be determined, $\Theta$.
Rather than continue with our undetermined
multipliers, the simplest way to proceed is
to reformulate the problem in terms of
the fidelity {\it F} with the constraints imposed.
We do this by specifying $N+2$ possible POM
elements corresponding to the required values
(0, $\pi$ and $\Theta$) of $\theta_k$:
\begin{equation}
\hat\pi_0=w_0
\left(
\begin{array}{cc}
2&0  \\
0&0
\end{array}
\right),
\label{pi_0}
\end{equation}
\begin{equation}
\hat\pi_\pi=w_\pi
\left(
\begin{array}{cc}
0&0  \\
0&2
\end{array}
\right),
\label{pi_pi}
\end{equation}
\begin{equation}
\hat\pi_l=w_l
\left(
\begin{array}{cc}
1+\cos {\Theta} & {\rm e}^{-i\phi_l}\sin {\Theta}  \\
{\rm e}^{i\phi_l}\sin {\Theta} & 1-\cos {\Theta}
\end{array}
\right),
\quad (l=1, ..., N).
\end{equation}
The fidelity is then
\begin{equation}
F={1\over2}
+{w_0\over2}\cos{\theta}(1+\cos{\theta})
+{w_\pi\over2}\cos{\theta}(1-\cos{\theta})
+W\left[  \cos^2{\theta}(1+\cos{\theta}\cos{\Theta})^2
              +{1\over4}\sin^4{\theta}(1-\cos^2{\Theta})
    \right]^{{1\over2}}.
\end{equation}
where $W=\sum_{l=1}^N  w_l$.  We can impose the
constraints, Eqs. (\ref{constraint1}) and (\ref{constraint2}),
in order to remove
$w_0$ and $w_\pi$ which leaves us with
\begin{equation}
F={1\over2}
+{1\over2}(1-W)\cos{\theta}
-{1\over2}W\cos^2{\theta}\cos{\Theta}
+{1\over2}W
    \left[  \cos^2{\theta}(1+\cos{\theta}\cos{\Theta})^2
              +{1\over4}\sin^4{\theta}(1-\cos^2{\Theta})
    \right]^{{1\over2}}.
\end{equation}
Extremising this fidelity to obtain the global maximum
value now corresponds to satisfying the conditions
\begin{equation}
{{\partial F}\over{\partial W}}={1\over2}
\left\{
-\cos{\theta}
-\cos^2{\theta}\cos{\Theta}
+ \left[  \cos^2{\theta}(1+\cos{\theta}\cos{\Theta})^2
              +{1\over4}\sin^4{\theta}(1-\cos^2{\Theta})
    \right]^{{1\over2}}
\right\}=0,
\label{dF_by_dW}
\end{equation}
\begin{eqnarray}
{{\partial F}\over{\partial \Theta}}
&=&-\sin{\Theta}{W\over2}
\left\{
-\cos^2{\theta}
+ \left[  \cos^2{\theta}(1+\cos{\theta}\cos{\Theta})^2
              +{1\over4}\sin^4{\theta}(1-\cos^2{\Theta})
    \right]^{-{1\over2}}
\left[  \cos^3{\theta}
+\cos{\Theta}(\cos^4{\theta}
              -{1\over4}\sin^4{\theta})
    \right]
\right\} \nonumber\\
&=&0.
\end{eqnarray}
The solution ${\rm sin}\Theta = 0$ corresponds to the
POM elements Eqs. (\ref{pi_0}) and (\ref{pi_pi}).
Combining the remaining
non-trivial
solution with (Eq. \ref{dF_by_dW}) leads to the appealingly simple
result that ${\rm cos}\Theta = 0$.  Hence the fidelity
has the form
\begin{equation}
F={1\over2}(1+\cos{\theta})
+{W\over2}
    \left[  (\cos^2{\theta}+{1\over4}\sin^4{\theta})^{1\over2}
              -\cos{\theta}
    \right].
\end{equation}
The maximum value that this can take clearly corresponds
to setting $W=1$ and hence the global maximum value
that the fidelity can take for symmetric qubit-states
is
\begin{equation}
F_{\rm max}=
1-{1\over4}\sin^2{\theta}.
\end{equation}
This takes its maximum value of unity for $\theta=0$.
This is reasonable as in this case all the signal states
correspond to $\ket{+}$ and unit fidelity can always
be achieved by choosing the single state $\ket{+}$
for retransmission.  The fidelity is a monotonically
decreasing function of $\theta$ and takes its
smallest value of $3/4$ for $\theta=\pi/2$
\cite{Fuchs_00}.

\begin{figure}[htb]
\centerline{\psfig{file=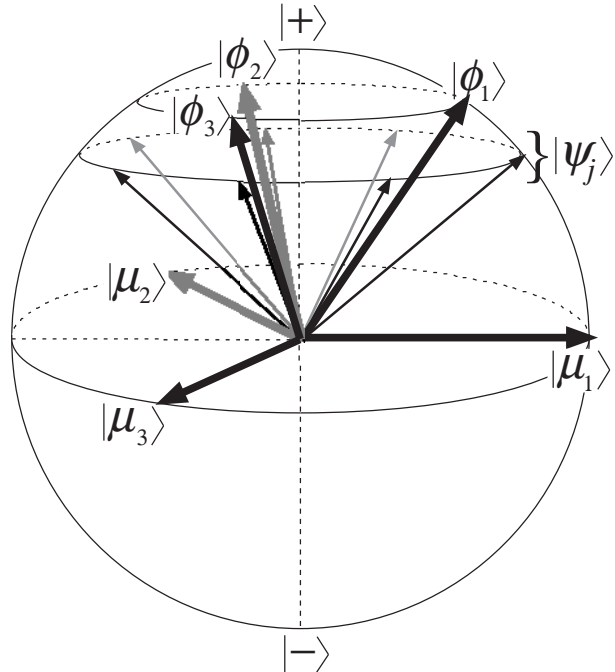,width=8cm}}
\caption{Optimal strategy that attains the maximum fidelity.
There are various possible
optimal measurement strategies as explained in the text.
In this figure, the measurement with three outputs $(M=3)$
and $\alpha=0$, and
the corresponding retransmission states are shown.
The retransmission states are positioned at the same longitude
($\phi$ coordinate) as the corresponding POM
elements but are further north than the original signal states.}
\label{POM}
\end{figure}

Having determined the maximum value of the fidelity,
we now turn our attention to the form of Bob's
strategy for realising this value.  The conditions
$\Theta=\pi/2$ and $W=1$ tell us that the optimal
POM will have $N$ elements of the form
\begin{equation}
\hat\pi_l=w_l
\left(
\begin{array}{cc}
1 & {\rm e}^{-i\phi_l} \\
{\rm e}^{i\phi_l} & 1
\end{array}
\right)
=w_l
    \left( \ket{+}
           + {\rm e}^{i\phi_l}\ket{-}
    \right)
    \left( \bra{+}
           + {\rm e}^{-i\phi_l}\bra{-}
    \right),
\label{pi_l}
\end{equation}
where the parameters $w_k$ and $\phi_k$ satisfy
the constraints
\begin{equation}
\sum_{l=1}^N w_l=1,
\quad (0\le w_l \le1)
\label{constraint1b}
\end{equation}
\begin{equation}
\sum_{l=1}^N w_l
{\rm e}^{i\phi_l}=0,
\label{constraint2b}
\end{equation}
corresponding to Eqs. (\ref{constraint1}) and (\ref{constraint2}) respectively.
The problem of maximising the fidelity does not
constrain the choice of POM any further than this
and so {\it any} POM with elements of the form
(Eq. \ref{pi_l}) and satisfying the conditions Eqs. (\ref{constraint1b})
and (\ref{constraint2b}) will maximise the fidelity.  Important
examples include the symmetric POM with elements
\begin{equation}
\hat\pi_l={1\over N}
\left(
\begin{array}{cc}
1 & {\rm exp}\left[-i(\alpha+{{2\pi l}\over{N}})\right] \\
{\rm exp}\left[i(\alpha+{{2\pi l}\over{N}})\right] & 1
\end{array}
\right),
\quad (N\ge2)
\label{pi_sym}
\end{equation}
where $\alpha$ is any desired phase.  We note that
the choice $N=2$ corresponds to a simple von
Neumann measurement.  Furthermore, setting
$N=M$ and $\alpha=0$ shows that the square-root
measurement, with POM elements Eq. (\ref{pi_mu})
can also maximise the fidelity.
An example of the states corresponding
the POM with elements Eq. (\ref{pi_sym}) is
depicted in Fig. \ref{POM}.
%%%%%%%%%%%%%%%%%%%%
%%%%%%%%%%%%%%%%%%%%

The retransmission states $\ket{\phi_l}$ that maximise
the fidelity correspond to the maximum-eigenvalue
eigenstates of the operator Eq. (\ref{O_k_enroute}) with
$\theta_k =\Theta=\pi/2$.
This operator is
\begin{equation}
\hat O_l={w_l\over2}
\left(
\begin{array}{cc}
1+\cos{\theta}
   &{1\over2}{\rm e}^{-i\phi_l}\sin^2{\theta}  \\
{1\over2}{\rm e}^{i\phi_l}\sin^2{\theta}
   &1-\cos{\theta}
\end{array}
\right)
\end{equation}
and the corresponding maximum eigenvalue is
\begin{equation}
\nu
_+(l)= w_l
\left(
1-{1\over4}\sin^2{\theta}
\right).
\end{equation}
Solving the for the associated eigenvector gives  the
required retransmission state associated with the
measurement outcome `$k$':
\begin{eqnarray}
\ket{\phi_l}&=&{1\over{\sqrt2}}
\left(
1+\cos^2{\theta}\right)^{-{1\over2}}
\left[
\left(
1+\cos{\theta}\right)\ket{+}
+{\rm e}^{i\phi_l}
\left(
1-\cos{\theta}\right)\ket{-}
\right] \nonumber\\
&=&
\cos{\left({\chi}\over{2}\right)}\ket{+}
+{\rm e}^{i\phi_l}\sin{\left({\chi}\over{2}\right)}\ket{-}.
\end{eqnarray}
These states are depicted on the Bloch sphere in
Fig. \ref{POM}.
They are positioned at the same longitude
($\phi$ coordinate) as the corresponding POM
elements but are further north than the original
signal states, having a lattitude $\pi/2 - \chi$
where
\begin{equation}
\cos{\chi}=
{2\cos{\theta}\over{1+\cos^2{\theta}}}.
\label{cos_chi}
\end{equation}

We can now summarise our strategies for obtaining
the maximum fidelity.  Any POM with elements
given by Eq. (\ref{pi_sym}) constitutes an optimum measurement.
These operators will form a POM if the conditions
Eqs. (\ref{constraint1b}) and (\ref{constraint2b}) are satisfied.  The 
fidelity will
be maximised if the retransmission state selected
on the basis of a the measurement outcome `$l$'
has the polar coordinates ($\chi, \phi_l$) on the
Bloch sphere, with $\chi$ given by Eq. (\ref{cos_chi}).

\subsection{Case 2: M=2}

\begin{figure}[htb]
\centerline{\epsfig{file=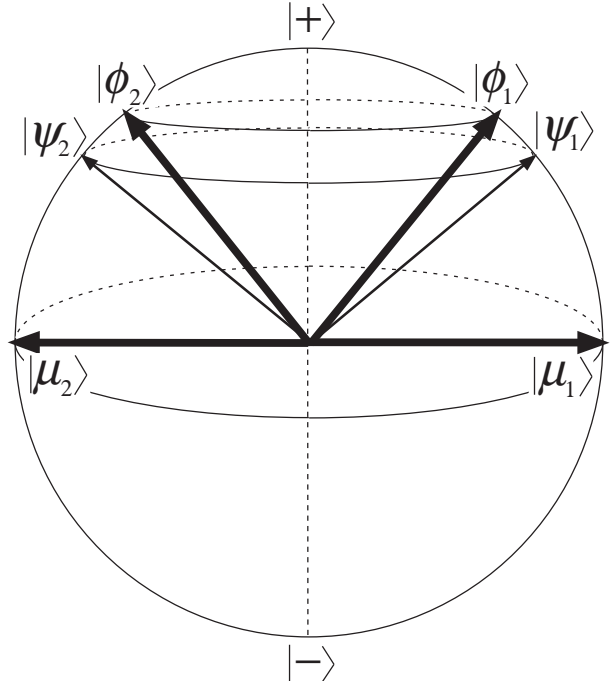,width=8cm}}
\caption{The case of two signal states.
In this case the optimal solution is unique, and comprises
a von Neumann measurement and the retransmission
of two nonorthogonal states.
The retransmission states are again further
north than the original signal states.
All the relevant states have the same longitude.}
\label{M=2}
\end{figure}

If $M=2$ then we have only two possible signal states:
\begin{equation}
\ket{\psi_j}= {\rm cos}\left({\theta\over2}\right)\ket{+}
                    \pm
                       {\rm sin}\left({\theta\over2}\right)\ket{-}.
\quad (0\le\theta\le{\pi\over2})
\label{general_psi_j_2}
\end{equation}
The representation of these states on the Bloch sphere
is depicted in Fig. \ref{M=2}.
%%%%%%%%%%%%%%%%%%%%
%%%%%%%%%%%%%%%%%%%%
The two states correspond to two
points at the same latitude in the northern hemisphere
with longitudes separated by $\pi$.  For these two states
the operator (\ref{O_k_gen}) becomes
\begin{equation}
\hat O_k={w_k\over2}
\left(
\begin{array}{cc}
(1+\cos{\theta})(1+\cos{\theta}\cos{\theta_k})
   &\sin^2{\theta}\sin {\theta_k}\cos{\phi_k}  \\
\sin^2{\theta}\sin {\theta_k}\cos{\phi_k}
   &(1-\cos{\theta})(1+\cos{\theta}\cos{\theta_k})
\end{array}
\right),
\label{O_k_two}
\end{equation}
the eigenvalues of which are
\begin{equation}
\nu
_\pm(\theta_k,\phi_k)={w_k\over2}
\left\{
1+\cos{\theta}\cos{\theta_k}
\pm
\left[
\cos^2{\theta}
(1+\cos{\theta}\cos{\theta_k})^2
+
\sin^4{\theta}\sin^2{\theta_k}\cos^2{\phi_k}
\right]^{1/2}
\right\}.
\end{equation}
The required greater of the two eigenvalues is clearly
maximised by setting $\phi_k=0$ or $\pi$.  Thus the strategy
that maximises the fidelity {\it must} comprise only POM
elements corresponding to states at the same longitudes
as the two signal states.

The maximisation of the fidelity
(\ref{max_fidelity})
follows the same lines as that for the case $M>2$ and
we will only present the main results.  The extremisation
of the fidelity subject to the constraints
(\ref{constraint1}) and (\ref{constraint2})
leads to the conclusion that the only possible values for
$\theta_k$ are $0,\pi$ and one other angle $\Theta$.
Repeating the extremisation with these possible
values for $\theta_k$ leads to the result that
$\Theta=\pi/2$ and that this is the value for which
the fidelity can attain its maximum value.  It follows
that the {\it unique} measurement strategy for maximising
the fidelity with the two possible signal states
(\ref{general_psi_j_2}) has the two POM elements
\begin{equation}
\hat\pi_k={1\over2}
\left(
\begin{array}{cc}
1 & (-1)^{k-1}  \\
(-1)^{k-1} & 1
\label{POM_2}
\end{array}
\right).
\end{equation}
This corresponds to a von Neumann measurement, the two
possible outcomes of which correspond to the two orthonormal
states
\begin{equation}
\ket{\mu_k}={1\over{\sqrt2}}\left(
\ket{+}+(-1)^{k-1}\ket{-}\right).
\end{equation}
This is the strategy that also minimises the error probability
if we associate the measurement outcome `$k$' with the
signal state $\ket{\psi_k}$.
The resulting maximum possible value for the fidelity  is
\begin{equation}
F_{\rm max}={1\over2}\left[
1+\left(\cos^2{\theta}+\sin^4{\theta}\right)^{1\over2}
\right].
\end{equation}
This takes its maximum value of unity for both $\theta=0$
and $\theta=\pi/2$.  This is reasonable as for $\theta=0$
all the signal states correspond to $\ket{+}$ and unit
fidelity can be achieved by simply retransmitting $\ket{+}$.
For $\theta=\pi/2$ the two signal states
(\ref{general_psi_j_2}) are orthogonal
and a von Neumann measurement
can determine the signal state with certainty.
The required retransmission state is then simply the
signal state.

The required retransmission states that achieve this maximum
fidelity are the eigenstates of the two operators
\begin{equation}
\hat O_k={1\over4}
\left(
\begin{array}{cc}
(1+\cos{\theta})
   &(-1)^{k-1}\sin^2{\theta}  \\
(-1)^{k-1}\sin^2{\theta}
   &(1-\cos{\theta})
\end{array}
\right),
\end{equation}
having the common greater eigenvalue
\begin{equation}
\nu_+={1\over4}\left[
1+\left(
\cos^2{\theta}+\sin^2{\theta}
\right)^{1\over2}
\right].
\end{equation}
These retransmission states are
\begin{eqnarray}
\ket{\phi_k}&=&{1\over{\sqrt2}}
\left(\cos^2{\theta}+\sin^4{\theta}\right)^{-{1\over4}}
\left[\left(
\cos^2{\theta}+\sin^4{\theta}\right)^{1\over2}
-\cos{\theta}\right]^{-{1\over2}}
\left\{\sin^2{\theta}\ket{+}+(-1)^k
\left[\left(\cos^2{\theta}+\sin^4{\theta}
\right)^{1\over2}-\cos{\theta}
\right]\ket{-}\right\} \nonumber\\
&=&
\cos{\left({\chi_2}\over{2}\right)}\ket{+}
+(-1)^k\sin{\left({\chi_2}\over{2}\right)}\ket{-}
\label{retransmit_2}
\end{eqnarray}
and are depicted on the Bloch sphere in Fig. \ref{M=2}.
They have the same longitude as the corresponding
POM elements but are again further north than the original
signal states having a latitude $\pi/2-\chi_2$ where
\begin{equation}
\cos{\chi_2}=\cos{\theta}
\left(\cos^2{\theta}+\sin^4{\theta}\right)^{-{1\over2}}.
\end{equation}
This angle also corresponds to a latitude that is
south of the optimum for the cases in  which $M>2$.

If there are only two possible signal states then the
maximum fidelity is achieved by means of the
{\it unique} strategy of performing the simple
von Neumann measurement corresponding to the POM
elements (\ref{POM_2}).  The required retransmission
states associated with the relevant measurement
outcomes have the form given in Eq.
(\ref{retransmit_2}).

\section{Conclusion}\label{Conclusion}

In a quantum communications channel the signal comprises
a known set of quantum states, each with a known
{\it a priori} probability for transmission.  The possibility of
selecting non-orthogonal states distinguishes the quantum
channel from its classical counterpart and leads to novel
technical possibilities including quantum key distribution
\cite{Phoenix_95}.  It also creates the interesting problem
for the receiver of having to select between a number of
possible detection strategies.  The decision will be informed
by the purpose for which the information retrieved is intended.
The strategy that minimises the error probability will have the
greatest chance of retrieving the number `$j$' associated with
the initial {\it classical} selection of the signal state.  As
such it is a measure of the quality of the measurement strategy
for retrieving this classical information.   The fidelity, however,
determines how well a state, selected on the basis of the
measurement outcome, will match the originally transmitted
signal state.  As such it depends on {\it both} the choice of
measurement strategy and the selection of the associated
`retransmission' states.  The fidelity measures the quality
of the measurement strategy for retrieving $\ket{\psi_j}$
rather than `$j$' and as such is a measure of the receiver's
ability to recover the {\it quantum} information in the
signal.

The fundamental difference between the error
probability and the fidelity may be illustrated with a
simple example. Suppose that the $M$ equiprobable
signal  qubit states are {\it all} of the form $\ket{+}$.
In this case, there is no measurement that can
decrease the error probability below the value
$1-1/M$ obtained by guessing the state.  Selecting
$\ket{+}$ as the only retransmission state, however,
gives the greatest possible fidelity of unity.

In this paper we have derived the maximum possible
fidelity for the symmetric qubit states defined in
Eq. (\ref{general_psi_j}).  This maximum value
depends on whether there are two or more than two
possible signals states.  For more than two signal
states there is a wide range of suitable measurement
strategies that can achieve the maximum fidelity.
This includes the unique strategy that minimises the
error probability.  For two signal states the only
strategy that can achieve the maximum fidelity is
that which minimises the error probability.  In
general, the required retransmission states
depend on the measurement outcome but
coincide with neither the signal states nor the elements
of the measurement POM.  It remains an open question
as to whether, for all possible states, the strategy that minimises the
error probability will always maximise the fidelity.
We will return to this question elsewhere.

\acknowledgements

We are grateful to Dr R. Willox and Prof. O. Hirota for helpful and
timely comments.  This work was supported in part by the
British Council.  SMB thanks the
Royal Society of Edinburgh and the Scottish Executive Education and
Lifelong Learning Department for the award of a Support Research
Fellowship.

%\end{multicols}


\begin{references}
\bibitem{foot1}
    There are, of course, more sophisticated strategies available to
    the eavesdroper than measurement and retransmission and the
    quality of these will be determined by other measures.
\bibitem{foot3}
    We note that we can also accommodate situations in which
    $\hat V^M = \hat I e^{i \alpha}$, for some phase $\alpha$, by
    replacing $\hat V$ with $\hat V' = \hat V e^{-i \alpha /M}$.
\bibitem{foot2}
    The reason for this is that the fidelity is linear in the POM elements
    and hence rank-two projectors can be included as a pair of rank-one
    projectors.
\end{references}
\end{document}